\documentclass[prl,twocolumn,showpacs,reprint,preprintnumbers,amsmath,amssymb]{revtex4}

\usepackage{bm}
\usepackage{graphicx}        
\usepackage{longtable}       
\usepackage{amsfonts}        
\usepackage{amsmath}         
\usepackage{amssymb}         
\usepackage{bm}              
\usepackage{color}           
\usepackage{dcolumn}         
\usepackage{epsfig}          
\usepackage{hyperref}        
\preprint{submitted to \PRL}

\newcommand{\JPCM}{J. Phys.: Condens. Matter }
\newcommand{\PRB}{Phys. Rev. B }
\newcommand{\PRL}{Phys. Rev. Lett. }
\newcommand{\RMP}{Rev. Mod. Phys. }

\begin{document}

\title{Stability of Weyl points in magnetic half-metallic Heusler compounds}

\author{Stanislav Chadov}\email{stanislav.chadov@cpfs.mpg.de}
\affiliation{Max-Planck-Institute for Chemical Physics of Solids,
             D-01187 Dresden, Germany}

\author{Shu-Chun Wu}
\affiliation{Max-Planck-Institute for Chemical Physics of Solids,
             D-01187 Dresden, Germany}

\author{Claudia Felser}
\affiliation{Max-Planck-Institute for Chemical Physics of Solids,
             D-01187 Dresden, Germany}

\author{Iosif Galanakis}\email{galanakis@upatras.gr}
\affiliation{Department of Materials Science, School of Natural
Sciences, University of Patras,  GR-26504 Patra, Greece}

\date{\today}

\begin{abstract}
We employ {\it ab-initio} fully-relativistic electronic structure
calculations to study the stability of the Weyl points in the
momentum space within the class of the half-metallic ferromagnetic
full Heusler materials, by focusing on Co$_2$TiAl as a
well-established prototype compound. Here we show that both the
number of the Weyl points together with their $k$-space
coordinates can be controlled by the orientation of the
magnetization. This alternative degree of freedom, which is absent
in other topological materials (e.g. in Weyl semimetals),
introduces novel functionalities, specific for the class of
half-metallic ferromagnets. Of special interest are Weyl points
which are preserved irrespectively of any arbitrary rotation of
the magnetization axis.
\end{abstract}

\pacs{75.30.-m, 75.10.Lp, 75.50.-y, 75.70.Tj}
\maketitle

The transfer of concepts between different disciplines and fields
of science has triggered the appearance and growth of new research
fields. One of the most striking recent examples is the notion of
topology in condensed matter physics and materials science.
Following the discovery of topological insulators
(see Refs. \cite{Ando} and \cite{Bansil} and references therein),
it was suggested theoretically that the band structures of
Na$_3$Bi and Cd$_3$As$_2$ show Dirac-like energy cones at the
Fermi surface \cite{Na3Bi,Cd3As2}. Theoretical predictions were
confirmed in 2014 and the compounds where named as Dirac
Semimetals \cite{Na3Bi-exp}. These findings initiated even more
intense interest on the topological effects and the role of the
breaking of symmetries was investigated \cite{Bernevig}.

In condensed matter physics the spin degeneracy at a general
$k$-point is protected by the coexistence of time-reversal
symmetry and inversion symmetry, the so-called Kramer degeneracy.
Breaking only one of the two symmetries breaks a Dirac point into
two Weyl points (WPs) of opposite chirality. Thus there are two
types of Weyl semimetals: (a) magnetic Weyl semimetals where
inversion symmetry is kept and (b) nonmagnetic noncentrosymmetric
ones where time-reversal symmetry is kept (for an extensive review
on Weyl semimetals see Ref.~\cite{Weng}). Two types of WP's can
occur: type-I, which are isolated points in the Brillouin zone and
type II representing a closed loop. There are also rare cases of
semimetals, like SrSi$_2$, where both the inversion and
time-reversal are broken leading to the formation of an exotic
double Weyl fermion \cite{Hasan2}.

\begin{figure*}
\includegraphics[scale=0.6,angle=270]{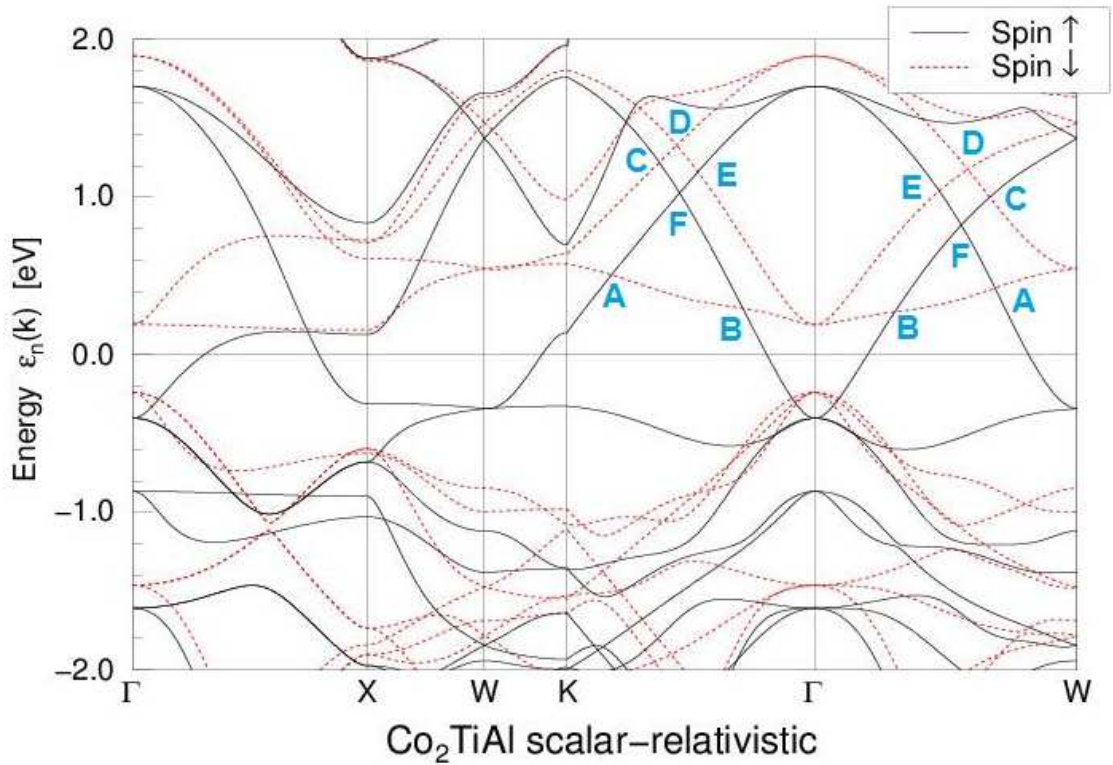}
\vspace*{-0.2cm} \caption{Calculated electronic band structure
 of Co$_2$TiAl using the scalar-relativistic formalism.  The zero energy
in the vertical axis is the Fermi level.  We have denoted the six
crossing points of interest with the symbols: A, B, C, D, E and F.
Note that all high symmetry-points are within the ${k_z=0}$ plane.}
\label{fig1}
\end{figure*}

The search for WPs in magnetic systems is a nontrivial task. First,
experimentally ARPES measurements of the magnetic Weyl materials
are very difficult due to the complex domain structure and thus
there is still no material definitely confirmed to be a magnetic
Weyl semimetal \cite{Weng}. Second, theoretically the search for
WPs in the 3D reciprocal space is rather complicated, since the
WPs might be general points of the Brillouin zone and not just
along high symmetry axes.  Similar manifolds might be formed also
by normal degeneracies along the high symmetric directions. In the
case of the magnetic systems such symmetry analysis requires the
Shubnikov type-II groups and becomes even more complicated.

Among the Heusler compounds \cite{Heusler,Heusler2}, the ones
being half-metallic ferromagnets have attracted considerable
attention due their potential applications in spintronics
\cite{Perspectives,book}. Their most striking characteristic is
the so-called Slater-Pauling behavior of the total spin magnetic
spin moment which scales linearly with the number of valence
electrons \cite{Galanakis,FelserRev}.  This large family provides
flexible possibilities to tune their electronic structure, e.g. by
shifting the Fermi energy with chemical composition. Due to the
very large number of Heusler compounds, it is natural to search for WPs either in the
nonmagnetic semi-Heuslers which crystalize in a
non-centrosymmetric lattice \cite{Binghai}, or among the regular
magnetic ones crystallizing in the L2$_1$ structure (space group
Fm$\overline{3}$m (225)). Their cubic crystalline structure
results into a low magnetocrystalline anisotropy, thus their
magnetization can be easily manipulated using  an external
magnetic field, both in single-crystalline as well as
polycrystalline samples. Since the L2$_1$ structure is
centrosymmetric, the appearance of magnetism breaks the
time-reversal symmetry leading possibly to the appearance of WPs.

Within 2016 three articles have appeared focusing on the WPs in
magnetic Heusler compounds. K\"ubler and Felser suggested that in
the case of Co$_2$MnAl the appearance of a large anomalous Hall
effect is linked to four WPs just above the Fermi level
\cite{Kubler}. Wang and collaborators studied several Co-based
Heusler compounds focusing especially on Co$_2$ZrSn and found that
when the magnetization is along the [110] direction there are at
least two WPs close to the Fermi level separated by a large
distance in the reciprocal space giving rise to well-defined Fermi
arcs \cite{Bernevig2}. Finally Chang and collaborators studied the
Co$_2$Ti(Si, Ge or Sn) compounds and found another two WPs in the
$K-\varGamma$ direction which are different than the ones
mentioned just above \cite{Hasan}. Ab-initio calculations in both
Refs. \cite{Bernevig2} and \cite{Hasan} revealed that the
different magnetization axis cannot be energetically distinguished
probably due to the very high symmetry of the lattice.  Thus the
important question rises concerning the stability of the observed
WPs with respect to the magnetization axis which can be rotated
due to external magnetic fields. The answer to this question is
primordial for applications since in most cases polycrystalline
films are used and thus an external magnetic field would oblige
the different domains to have their magnetization along different
crystallographic axis and thus the number of interesting band
crossings might be substantially reduced. The most interesting
band crossings would be those which survive upon arbitrary
rotation of the magnetization.

To study them, we consider a typical representative of the
Co$_2$-based Heusler family, namely Co$_2$TiAl, which has been
well established experimentally and theoretically as a good
half-metallic ferromagnet obeying the Slater-Pauling rule; it has
25 valence electrons per formula unit with a total spin magnetic
moment of 1 $\mu_\mathrm{B}$ per formula unit \cite{Galanakis}.
We have also studied  Co$_2$TiSi and Co$_2$VAl which have 26
valence electrons, and results and conclusions were also similar
for them with the only noticeable difference being that the WPs
were lower in energy due to the extra electron as expected; the
Fermi level in Heuslers can be varied within a rather wide range
by changing the chemical composition \cite{book}.  In all cases
we have used the experimental lattice constants
\cite{Buschow,Barth}. To perform the electronic structure
calculations, we employed the full-potential nonorthogonal
local-orbital minimum-basis band structure scheme (FPLO)
\cite{FPLO} within the generalized gradient approximation (GGA)~\cite{GGA}. We
performed first scalar-relativistic and then
fully-relativistic calculations where the Dirac-like analogue to
Kohn-Sham equations are solved \cite{FPLO}. As the quantization axis
we have considered three cases: [001], [110] and
[111]. During the self-consistency, for the Brillouine zone integration a Monkhorst-Pack 30$\times$30$\times$30
grid has been used \cite{Monkhorst}. Total energies have been
converged with an accuracy of 10$^{-8}$ in Hartree. Band structure
plots have been plotted within the ${k_z=0}$ plane for all cases
considering a very dense division of 500 \textbf{k}-points between
two consecutive high symmetry points.

\begin{figure}
\includegraphics[scale=0.33,angle=270]{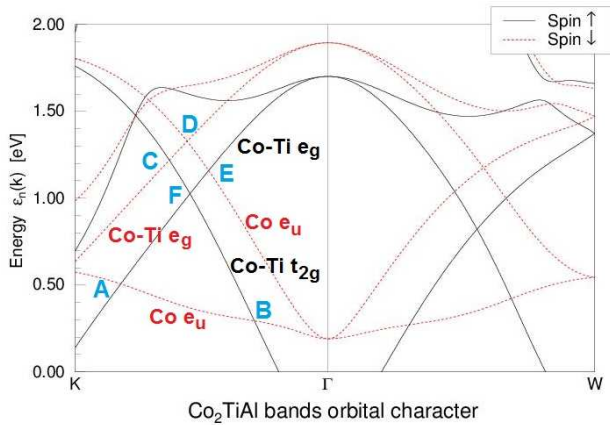}
\vspace*{-0.2cm} \caption{Orbital character of the bands which
cross at the six points along the $K-\varGamma-W$ direction shown
in Fig. \ref{fig1}.} \label{fig2}
\end{figure}

We will start our discussion from the scalar-relativistic band
structure obtained for Co$_2$TiAl and presented in Fig.~\ref{fig1}. In this case we
can decompose the bands in pure majority- and minority-spin
character. The minority-spin band structure presents a direct gap
at the $\varGamma$ point of about 0.5\,eV typical for half-metallic
Heusler compounds and the total spin magnetic moment of 1\,$\mu_\mathrm{B}$ in agreement with the Slater-Pauling rule~\cite{Galanakis}.
All high-symmetry points in Fig.~\ref{fig1} have
been chosen within the ${k_z=0}$ plane perpendicular
to the $z$-axis; $\varGamma$, $X$, $W$ and $K$ coordinates in the
reciprocal space are $[0\:0\:0]$, $[1\:0\:0]$,
$[1\:\frac{1}{2}\:0]$ and $[\frac{3}{4}\:\frac{3}{4}\:0]$,
respectively, in $\frac{2\pi}{a}$ units where $a$ is the lattice
constant. Due to the existence of three transition metal atoms per
formula unit, there is a very large umber of $d$ bands around the
Fermi level in a narrow energy region (for the character of the
bands see Ref. \cite{Galanakis}). As a result several crossings
which can be considered as potential WPs occur not only along the
high symmetry lines shown in Fig. \ref{fig1} but in the whole
Brillouin zone.

\begin{figure}
\begin{center}
\includegraphics[scale=0.33,angle=270]{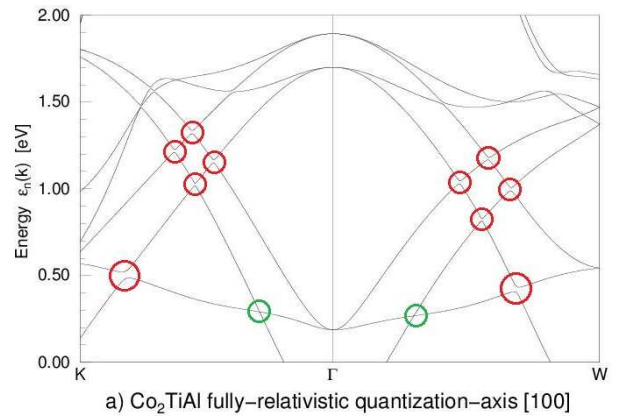}
\includegraphics[scale=0.33,angle=270]{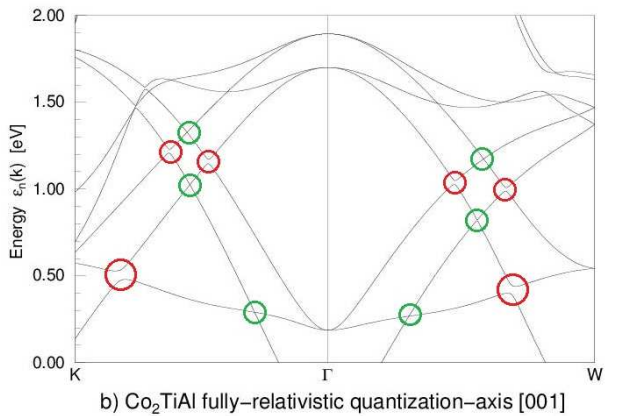}
\end{center}
\vspace*{-0.2cm} \caption{(a) Calculated electronic band structure
 of Co$_2$TiAl using the fully-relativistic formalism and setting
the quantization axis to be the [100]. (b) same with [001] as the
quantization axis. In both cases the zero energy in the vertical
axis is the Fermi level. We have encircled with red colors the
crossing points where the degeneracy is lifted and no crossing
occurs anymore, and with green color the ones where the crossing
is preserved. } \label{fig3}
\end{figure}

In order to make our study possible we have focused on six
crossing points along the $K-\varGamma$ direction which have their
exact symmetry analog in the $\varGamma-W$ direction as shown in
Fig. \ref{fig1}. The $\varGamma$-$K$ direction has been proposed
to contain the WPs according to Refs.~\cite{Bernevig2} and
\cite{Hasan}. The energy window under study which contains the
relevant WPs is between the Fermi level and 2 eV above it
(${0<E-E_{\rm F}<2}$\,eV) as shown in Fig.~\ref{fig1}.
 A detailed analysis of the band
structure using the so-called fat band scheme~\cite{CrSe} revealed
the character of the bands which cross and we present in Fig.
\ref{fig2} the bands orbital character as deduced using the fat
band scheme.  The minority-spin band at the A and B points, which
have been proposed to be WPs in Ref. \cite{Hasan}, belongs to the
double degenerate $e_u$ bands which have exclusively their weight
at the Co atoms (see Ref. \cite{Galanakis} for an extended
discussion of the character of the bands). This band at the point
B crosses a majority-spin band made up of states belonging the
triple-degenerate antibonding $t_{2g}$ states with most of their
weight a the Ti atom, and at point A crosses a majority-spin band
 belonging to the double degenerate antibonding
$e_g$ states again with most of their weight at the Ti atoms. The
majority-spin $t_{2g}$ band crosses after the B point the
antibonding majority-spin $e_g$ band at point F and the
minority-spin  $e_g$ band at point C. The other minority-spin
$e_u$ state discussed in the beginning crosses also these two
states at points E and D. Thus all six points arise due to
crossing of different bands.

\begin{figure}
\includegraphics[width=\columnwidth]{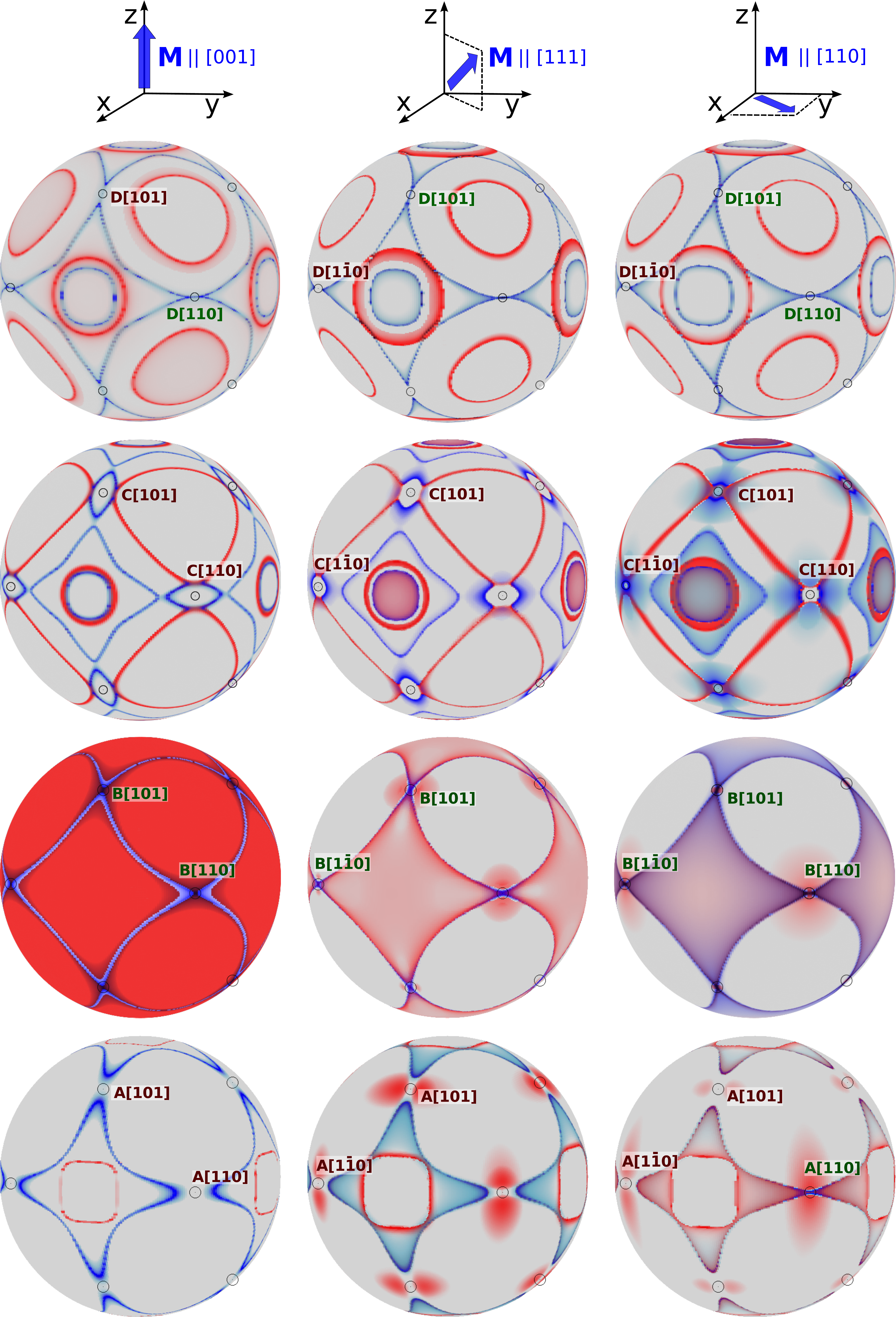}
\caption{Fully relativistic band structures of Co$_2$TiAl computed
along
  $K$-$\varGamma$-$K'$ $k$-path ($\varGamma$-$K$ and $\varGamma$-$K'$ directions are
  denoted explicitly) for three orientations of magnetization
  ${\bf M}$: [001],[111] and [110]. (${E,k}$)-points of
  interest (of A, B, C and D type) are emphasized by green or red circles depending on their
  crossing or anticrossing character, respectively. Minority- and
  majority-spin states are colored as red and blue, respectively.
   Below we mark the corresponding nonequivalent points of each type by
   plotting the spectral weight on $\varGamma$-centered spheres in 3D $k$-space (green
  letters mark crossing, red  - anticrossing characters). \label{fig4}}
\end{figure}

As a first step using the fully-relativistic formalism we have
computed the band structure choosing as quantization-axis the
[100] and [001] simulating this way the effect of the external
magnetic field on the direction of the magnetization, and we
present our results in Fig.~\ref{fig3} focusing above the Fermi
level and in the ${K-\varGamma-W}$ direction. To facilitate the
discussion we have encircled with red color the cases where the
degeneracy is lost and there is no crossing anymore and with green
when we the crossing is preserved. In the Dirac relativistic
formalism spin is no more a good quantum number and thus we cannot
project our band structures on spin. When the quantization axis is
along the [100] direction then only the $M_x$ mirror plane which
is normal to the [100] direction and the fourfold $C_{4x}$
rotational axis survive. As a result in the ${k_z=0}$ plane only
the WP at points B is preserved. The ${k_y=0}$ is equivalent to
${k_z=0}$ plane and the question is what happens in the ${k_x=0}$
plane normal to the [100] axis. To answer this question we have
considered as quantization axis the [001] and now the ${k_z=0}$
plane in the lower panel of Fig.~\ref{fig3} corresponds to the
${k_x=0}$ plane in the case of the [100] axis. We can see that now
at the points D and F the degeneracy is preserved. Thus in
magnetic materials the situation is much more complicated than the
one expected for magnetic semimetals~\cite{Weng}. We have
performed fully-relativistic calculations also considering the
[110], [011] and [111] as the quantization axis.  Now the symmetry
operations are different in each case, e.g. in the [110] case the
symmetry operations -except the inversion symmetry- which are
preserved are the mirror plane $M_{xy}$ which is normal to the
[110] axis and the $C_2^{[110]}$ rotation. Presented band
structures suggest that different crossing points are conserved in
each case.

To further analyze the behavior of the crossing points we focused
on the ones  marked as A, B, C and D in Fig.~\ref{fig1} since F is
equivalent to D and E  to C. In this case there are 12 equivalent
$\varGamma$-$K$ directions allowed by the Fm$\overline{3}$m
space-group. In order to distinguish between spin states in the
fully-relativistic regime, we have employed
the Korringa-Kohn-Rostoker (KKR) electronic structure method
employing relativistic spin-projection operators~\cite{KKR}. For the usual spin-polarized case without
the spin-orbit coupling KKR yields a similar band structure to
the FPLO one, presented in Fig. \ref{fig1}. In order to track the
appropriate changes induced by the spin-orbit coupling with
respect to the simply spin-polarized ones, we will simultaneously depict all 12
points of each type (A, B, C and D) on the same
$\varGamma$-centered sphere in the 3D $k$-space, by tuning its
radius precisely to host the given point type (and simultaneously
tuning the corresponding energy). Such representation allows
to compare their character together with its evaluation by changing the
magnetization orientation (see Fig.~\ref{fig4}).

In Fig.~\ref{fig4} we consider three distinct orientations of the
magnetization: ${\vec M\parallel[001]}$, [111] and [110]. We also
decompose the spectral density into the majority- and
minority-spin (using red and blue colors for the spectral density,
respectively). By considering, for example, the sphere containing
the point A (A-sphere), we see that it has an  anticrossing
character for all $k$-directions and magnetization orientations
except for $\vec M\parallel$[110], where it keeps crossing along
$\vec k\parallel$[110] (and along the  equivalent
[$\bar{1}\bar{1}$0] which is just hidden behind). One partially
observes this also in the corresponding $E(k)$ plots (shown only
for two nonequivalent directions): encircling of the A-points is
shown in red everywhere (anticrossing) except of $\vec
k\parallel$[110] for $\vec M\parallel$[110] case (green). This
means that by taking the polycrystalline case, the overall
contribution of the A-crossing into the transport properties
(provided it is tuned to the Fermi energy) will be drastically
reduced. The same situation occurs for the C-type, which keeps
crossing  only along $\vec k\parallel$[1$\bar{1}$0] (and
[$\bar{1}$10]) for $\vec M\parallel$[110] case. The intermediate
situation occurs for the D-type which has a pure minority-spin
character: for $\vec M\parallel$[001] it gives crossing characters
in four xy-directions: [110], [$\bar{1}$$\bar{1}$0], [1$\bar{1}$0]
and [$\bar{1}$10], six crossings in case of $\vec
M\parallel$[111], and six - in case of $\vec M\parallel$[110]. In
contrast to A,C, and D, the B-point exhibits the most universal
behavior: it preserves the crossing character in all
$k$-directions for all magnetization orientations. This indicates,
that by tuning the Fermi energy to the B-point, its efficient
overall transport response might be achieved even in the
polycrystalline case.

 Employing ab-initio calculations within the fully-relativistic
Dirac-formalism we studied the stability of Weyl points in
magnetic half-metallic Heusler compounds using Co$_2$TiAl as a
prototype. Our results suggest that the preservation of the
crossing points observed in the scalar-relativistic band structure
strongly depends on the orientation of the quantization axis due
to the broken symmetry operations. Moreover due to symmetry
reasons the band structure within the ${k_x=0}$, ${k_y=0}$ and ${k_z=0}$
plane differs for a given quantization axis making the
identification of the Weyl points which are preserved even more
complex. In the case of Co$_2$TiAl we found that there is a
crossing which is invariant to the rotation of the magnetization
axis and thus it is a real Weyl point.

Our results clearly show that external magnetic fields, which pin
the quantization axis in magnetic half-metals, affect the number
and the momentum space location of Weyl points offering novel
possibilities, not present in conventional Weyl semimetals,  in
the field of topological materials.

Authors acknowledge fruitful discussion with G.~H.~Fecher.
One of the authors (IG) gratefully acknowledges the
financial support by the Max-Planck Gesellschaft.

\end{document}